# *In silico* bioactivity prediction of proteins interacting with graphene-based nanomaterials guides rational design of biosensor

*Jing Ye[a,1], Minzhi Fan[a,1], Xiaoyu Zhang[a], Shasha Lu[b], Mengyao Chai[a], Yunshan Zhang[a], Xiaoyu Zhao[a,c], Shuang Li[d], Diming Zhang[a,\*]*

[a] Research Center for Intelligent Sensing Systems, Zhejiang Laboratory, Hangzhou, 311121, China

[b] School of Materials Science and Engineering, Suzhou University of Science and Technology, Suzhou 215009, China

[c] College of Materials and Environmental Engineering, Hangzhou Dianzi University, Hangzhou, 310018, China.

[d] Academy of Medical Engineering and Translational Medicine, Tianjin University, Tianjin, 300072, China

*Corresponding author: Diming Zhang (zhangdm@zhejianglab.edu.cn)

[1]These authors contributed equally.

**ABSTRAT**

Graphene-based nanomaterials have attracted significant attention for their potentials in biomedical and biotechnology applications in recent years, owing to the outstanding physical and chemical properties. However, the interaction mechanism and impact on biological activity of macro/micro biomolecules still require more concerns and further research in order to enhance their applicability in biosensors, etc. Herein, an integrated method has been developed to predict the protein bioactivity performance when interacting with nanomaterials for protein-based biosensor. Molecular dynamics simulation and molecular docking technique were consolidated to investigate several nanomaterials: C60 fullerene, single-walled carbon nanotube, pristine graphene and graphene oxide, and their effect when interacting with protein. The adsorption behavior, secondary structure changes and protein bioactivity changes were simulated, and the results of protein activity simulation were verified in combination with atomic force spectrum, circular dichroism spectrum fluorescence and electrochemical experiments. The best quantification alignment between bioactivity obtained by simulation and experiment measurements was further explored. The two proteins, RNase A and Exonuclease III, were regarded as analysis model for the proof of concept, and the prediction accuracy of protein bioactivty could reach up to 0.98. The study shows an easy-to-operate and systematic approach to predict the effects of graphene-based nanomaterials on protein bioactivity, which holds guiding significance for the design of protein-related biosensors. In addition, the proposed prediction model is not limited to carbon-based nanomaterials and can be extended to other types of nanomaterials. This facilitates the rapid, simple, and low-cost selection of efficient and biosafe nanomaterials candidates for protein-related applications in biosensing and biomedical systems.

# 1. Introduction

In recent years, graphene-based nanomaterials (GBNMs) have received substantial attention across a wide range of disciplines, including biotechnology (Wang et al. 2011), life science (Bianco et al. 2008) and environmental science (Cui et al. 2022) due to their remarkable properties of large surface ratio, easy modification and functionalization. Specifically, GBNMs span wide applications in the realm of protein-based biosensors, notably in ultraviolet (Yüce and Kurt 2017), fluorescence (Wen et al. 2015), electrochemical (Kour et al. 2020), and Raman scattering spectroscopy fields (Liang et al. 2021), because the notable capability for the non-covalent adsorption of a substantial quantity of bio-sensitive constituents, such as proteins and peptides (Hampitak et al. 2020). GBNMs have shown the advantages of enhanced signal response and high sensitivity in a variety of sensing applications (Cheng et al. 2017; Peña-Bahamonde et al. 2018; Niazi et al. 2023; Yang et al. 2023), for example, graphene-based field-effect transistors that are ultrasensitive to surface state changes and are widely used in biomedical diagnosis including SARS-CoV-2 detection (Unal et al. 2021). It is worth noting that in such applications, macro/micro biomolecules like proteins directly interact with materials, may lead to protein conformation distortion, even denaturation and losing their structural activities, which can be detrimental to functions of biomolecular detection (Badhe et al. 2021). While coupling reagents such as 1-pyrenebutanoic acid succinimidyl ester can mitigate protein inactivation by facilitating covalent modifications, their use undoubtedly introduces additional complexity to the sensor construction process. Therefore, gaining more insight of the interplay between GBNMs and biomolecules will contribute to identifying protein-compatible GBNMs the advancement of biosensor performance.

Various experimental tools and methodologies including circular dichroism (CD) (Girmatsion et al. 2022; Miles et al. 2021), dynamic light scattering (DLS) (Shrivastava et al. 2023), size exclusion chromatography (SEC) (Yüce and Kurt 2017; Leong et al. 2022), and spectroscopy technologies (Yang et al. 2022), has been used to examine the influence of non-covalent adsorption on protein structure of protein-nanomaterial complex. In view of the close relationship between protein structure and function, the adsorption process of proteins on the interface of nanomaterials is much more complicated than that of ordinary molecular systems (Lee et al. 2023). Conventional experimental methods are not only time-consuming, costly, complex to operation (Verma et al. 2022) but also difficult to study in depth the interaction between protein and material surface. It is of great theoretical and practical value to obtain the structural information of proteins and study the changes of structural properties of proteins during adsorption from the molecular level.

As the computer simulation method closest to experimental conditions, molecular dynamics simulation and molecular docking can predict and explain the interaction mechanism between molecules visually, and thus providing a beneficial supplement for the experimental research on the adsorption of various proteins on the surface of biological materials (Zeng et al. 2023). Therefore, molecular dynamics simulation and molecular docking techniques has been utilized to study the interactions between protein and nanomaterials, offering high spatial resolution insights into protein orientation and conformation (Hirano and Kameda 2021; Badhe et al. 2021). For the simulation between protein and nanomaterials, the characterization of protein behavior when interacting with and absorbed onto GBNMs can be delineated by three key parameters: surface-protein structure, surface curvature and affinity (Zuo et al. 2011). Some researches demonstrated that conformation transitions of peptides and proteins can be controlled by residue mutation, enhancing the understanding of mechanisms of interaction manner between protein and nanomaterials (Lu et al. 2019). In addition, investigating the affinity hot spots between proteins and carbon nanomaterials with surface curvature profile can help predict the stability of the entire structure, further explored as the nanoscale platform for drug delivery (Li et al. 2022). While the majority of existing simulations have predominantly focused on the comprehensive examinations of global protein structural alterations. However, there remains a dearth of systematic and direct simulation tactics regarding the activity or functionality changes of proteins subsequent to their interaction with nanomaterials.

In response to the aforementioned issue, we integrated molecular dynamics simulation and molecular docking to study the effects of non-covalent adsorption of graphene-based nanomaterials on both protein structure and protein bioactivity, and verify it with experiments herein (Fig. 1). In the study, RNase A and Exonuclease III (ExoIII) were selected as model proteins, and a general simulation method was proposed to investigate the interaction behavior between protein and several types of GBNMs (Fig. 1a): zero-dimensional fullerene C60 (C60), one-dimensional single-wall carbon nanotube (SWCNT), two-dimensional pristine graphene (PG) and graphene oxide (GO). Initially molecular dynamics simulation (Fig. 1b) was applied to examine the secondary and tertiary structural change of proteins, employing the material-compatible force field to improve the accuracy of the structure prediction, alongside consideration of favorable surface curvature. It is worth mentioning that the smaller size and structural properties of C60 nanoparticles significantly influence the interaction with protein, contrasting with other three GBNMs (Verma et al. 2022). All-atom simulation was applied for PG, GO and SWCNT while a coarse-grained simulation was utilized for C60 to simplify parameterization and highlight the key interactions between C60 and protein. Additionally, detailed biophysical properties of protein were further analyzed for accurate molecular docking, which provides insights into affinities between protein and corresponding ligands (Fig.1c) in purpose of studying the bioactivity of proteins before and after interacting with nanomaterials. Moreover, various snapshots of simulation were evaluated for the most accurate bioactivity prediction and correct quantification. Subsequently, experimental tools such as circular dichroism spectroscopy, atomic force microscopy, fluorescence spectroscopy and electrochemical detection was employed to probe alterations in the activity of proteins that whether they still maintain their functions upon their interaction with carbon-based nanomaterials and validated our simulation results. This study pioneered an uncomplicated and systematic approach to evaluate and measure the impact of nanomaterials on protein activity, achieving a precision level of 0.98. This has substantial implications for shaping the development of biosensors related to proteins. Moreover, the applicability of this method extends beyond carbon-based nanomaterials, encompassing the exploration of interactions between various proteins and nanomaterials. Consequently, it allows the identification of more effective and biologically safer nanomaterials for a range of protein-related applications, considering factors such as biocompatibility and efficacy.

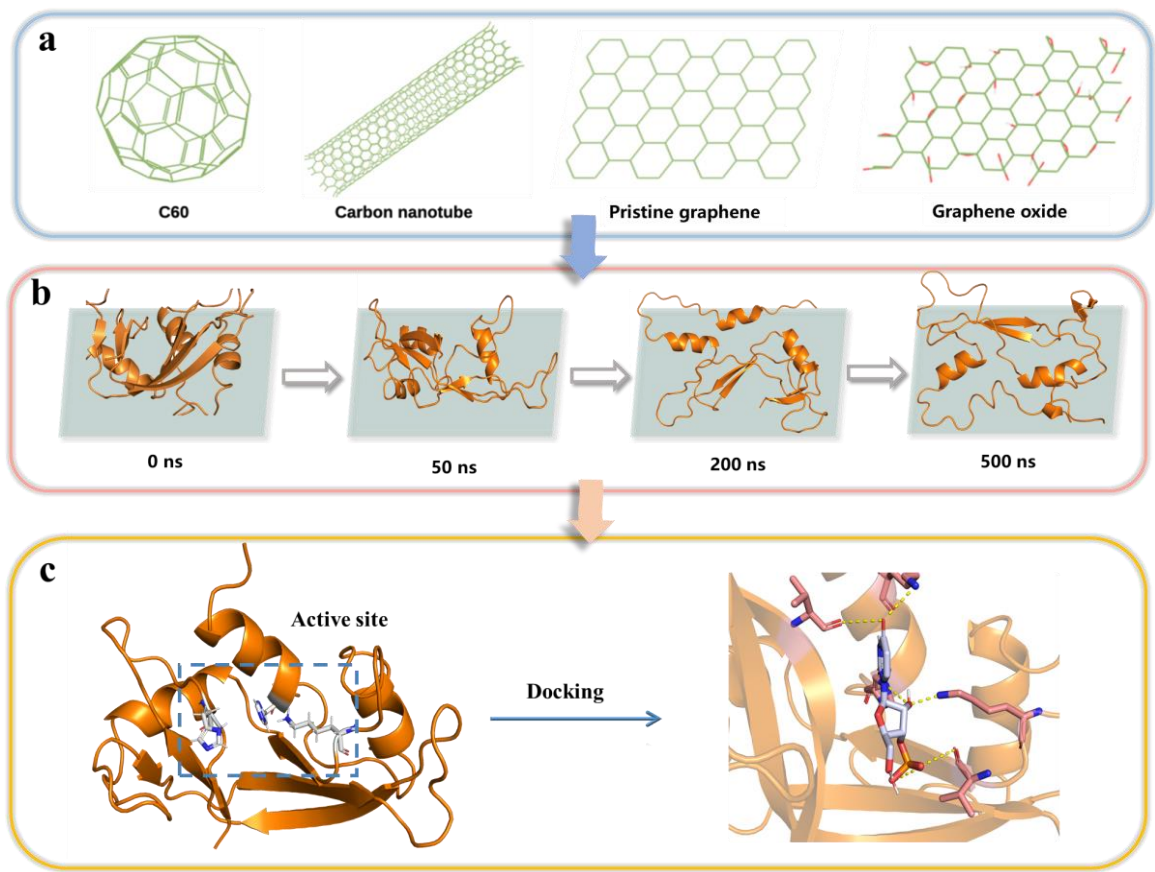

**Fig. 1. Schematic illustration of *in silico* prediction. (a)** Structure diagram of four GBNMs. **(b)** The conformational change process of protein to nanomaterial by molecular dynamic simulation. **(c)** The bioactivity simulation of protein by molecular docking.

## 2. Methods

*2.1 Reagents and apparatus*

RNase A (10 mg/ml) was obtained from Beyotime Biotechnology (Shanghai, China). PG and GO provided by chemical vaporization was purchased from XFNANO (Nanjing, China). SCNTs (2 mg/ml) and C60 was obtained from Wow Material (Suzhou, China). Methylene blue (MB) was purchased from Aladdin (Shanghai, China). $NaH_2PO_4$, $Na_2HPO_4$ and KCl were bought from Sinopharm (Shanghai, China). Ultra GelRed dye was bought from Vazyme Biotech (Nanjing, China). Ultrapure water (18.25 MΩ·cm, Milli-Q, Millipore) was used in all experiments. ExoIII and all the oligonucleotides (Table S1) were purchased from Sangon Biotech (Shanghai, China). The oligonucleotides powder is dissolved in pure water (100 μM), and dissolved with PBS buffer (10 mM $NaH_2PO4/Na_2HPO4$, pH 7.4) to the desired concentration for used. In order to make sure efficient hybridization between DNAs, the dissolved DNA solutions were firstly heated at 95°C for 5 min, then slowly cooled to room temperature and stored at 4°C for further use.

Circular dichroism (CD) spectroscopy was measure by Chirascan V100 (Applied Photophysics), with quartz cuvette of 0.5 cm path length in the far UV (180-260 nm). At least three full wavelength scans with bandwidth 1 nm/0.5 s were taken for each sample, and the spectra were averaged. The atomic force microscopy (AFM) of nanomaterial samples were inspected in situ using Dimension Icon ScanAsyst (Bruker, US), the images were obtained in tapping mode to analyze the morphologies and conformation of graphene, graphene oxide and carbon nanotube. Particle size analysis was carried out by nano particle analyzer (Malvern Panalytical Limited, UK), samples were diluted with distilled water and measured using quartz cuvette (200 μL) and cleaned the curvette every time after measurements with water, ethanol and water separately. Equilibrium time was set to 2 min and 3 trials were taken for each sample. The fluorescence emission spectra were performed on F-7100 Fluorescence Spectrophotometer, Hitachi High-Technology Co., Ltd. (Tokyo, Japan). The sample cell was a 200 μL quartz cuvette. The fluorescence intensities at 521 nm were monitored upon excitation at 488 nm. The slits for both of excitation and emission were set at 10 nm. The electrochemical experiments were conducted with Chenhua CHI 660E instrument (China) at 25 °C, employing a three-electrode system. The system consisted of a glassy carbon electrode (GCE, Φ=3 mm) as the working electrode, a platinum wire as the counter electrode and an Ag/AgCl as the reference electrode. Differential pulse voltammetry (DPV) was measured with the potential scan of -0.4 to 0 V and scan rate of 0.05 V/s.

2.2 Molecular Dynamics Simulation

3D structure of RNase A was obtained from RCSB PDB Protein Data Bank (1DFJ), all atom simulations were carried out on protein and different GBNMs using Amber 22 (Case et al. 2022) molecular dynamic package with ff19SB force field for protein, modified GAFF2 force field including GOPY (Muraru et al. 2020) for PG and GO, and BuildConstruct (Minoia et al. 2012) for SWCNT. Detailed force field was in supplementary materials. 3D structure of C60 was also obtained from RCSB PDB Protein Data Bank (60C). The topology and coordinate parameters of the system were generated by tLEaP and antechamber (Wang et al. 2006) modules of Amber22. All the molecular dynamics simulations were carried out in implicit solvent model.

A series of energy minimization steps were performed as follows: in the first stage, the system was minimized for 500 cycles of steepest descent followed by 1000 cycles of conjugate gradient with 50 kcal/mol/Å2 harmonic force constant restraints on the protein. Then the entire system was minimized for 1000 iterations with 50 kcal/mol/ Å2 restraints on all atoms. Heating was performed for 100 ps from 0K to 300K using Langevin thermostat dynamics with the collision frequency 2 ps$^{-1}$. Then the system was relaxed with backbone restraints at 2 kcal/mol/ Å2 at 300K for 200 ps and equilibrated freely for 1 ns in the NVT ensemble, maintaining constant pressure at 1 bar using Berendsen pressure scaling algorithm. A production run with no constraints was performed after for in total 500 ns for absorption process analysis. The long-range electrostatic interactions were treated with the particle-mesh Ewald method with a non-bonding cut-off distance of 10 Å and SHAKE algorithm was employed to constrain all hydrogen atoms. Finally, the simulation results (Fig. 1b) are obtained. Each nanomaterial simulation has been produced with three replicas, and for data analysis, trajectories of production run were extracted every 100 frames of total 2000 frames.

2.3 Molecular Docking

Molecular docking (Fig. 1c) was carried out by Gnina (McNutt et al. 2021). Molecular docking was performed twice for each protein with different ligands: the first structure was originally obtained from RCSB PDB Protein Data Bank and Protein Data bank in Europe; the second structure used was extracted from molecular dynamics simulation as the resulting structure after interacting with nanomaterials. In the docking process, ligands were set as fully flexible and rotatable while receptors were set as rigid. The grid box was restricted at the enzyme active site with a size of 1000 Å$^3$ and general_default2018_2 model was selected for docking and affinity calculation. Exhaustiveness was set to 64 and 10 poses of ligands were simulated.

2.4 Protein activity detection

RNase A (1 ng/μL) was placed to C60, SWCNT, PG and GO (5 μg/ml) solution at 25 °C for 30 min to obtain four RNase A-GBNMs bioconjugates, respectively. An equal amount of RNA1 (100 nM) was added to four RNase A-GBNMs bioconjugates, respectively, and reacted for 30 minutes to facilitate RNA hydrolysis (solution a). The mixture of DNA1 and DNA2 was firstly heated at 95 °C for 5 min, then slowly cooled to room temperature for further use. ExoIII (2 U/μL) was placed to C60, SWCNT, PG and GO (5 μg/ml) solution at 25 °C for 30 min to obtain four ExoIII-GBNMs bioconjugates, respectively. The DNA1/DNA2 duplex (100 nM) was added and reacted for another 30 min. After reaction, the solution was heated at 70 °C for 30 min to deactivate the enzyme (solution a). The mixture of helper DNA (40 nM), 6-carboxyfluorescein functioned hairpin DNA 1 (FAM-H1,100 nM), 6-carboxyfluorescein functioned hairpin DNA 2 (FAM-H2, 100 nM) were added successively to different solution a, and incubated at room temperature for 90 min to initiate hybridization chain reaction. The final solutions were diluted for fluorescence detection. The mixture of helper DNA (40 nM), hairpin DNA 1 (H1,100 nM), hairpin DNA 2 (100 nM) were added into various solution a at room temperature. After 90 min, methylene blue (final concentration is 10 uM) was added and reacted at room temperature for 30 min. The solution was used directly for electrochemical testing. Under the same conditions, the sample without GBNMs was regarded as a control sample.

## 3. Results and discussions

*3.1 Absorption behavior simulations*

Two proteins, RNase A and ExoIII were chosen as model proteins for this study. RNase A and ExoIII are biological enzymes capable of hydrolyzing RNA and DNA, respectively, offering a significant advantage for the subsequent validation of simulations through straightforward experimental techniques such as electrochemistry and fluorescence. The intrinsic structures of RNase A and ExoIII exhibit marked differences in their α-helices and β-strands, representing two proteins with distinct secondary structures. This diversity is valuable for the comprehensive nature of this study. There are pronounced differences in the critical functional sites between RNase A and ExoIII. This distinction highlights two types of functional protein architectures, providing some insight and reference value for future research. Consequently, these proteins are deemed representative and valuable for exploring the nuances of protein structure and function. Moreover, this study primarily focused on four GBNMs (C60, SWCNT, PG and GO), which share similarity that a single layer of carbon atoms arranged in a hexagonal lattice. By modifying PG structure such as rolling up or attaching functional groups to its surface, GO, SWCNT and C60 were created and modeled.

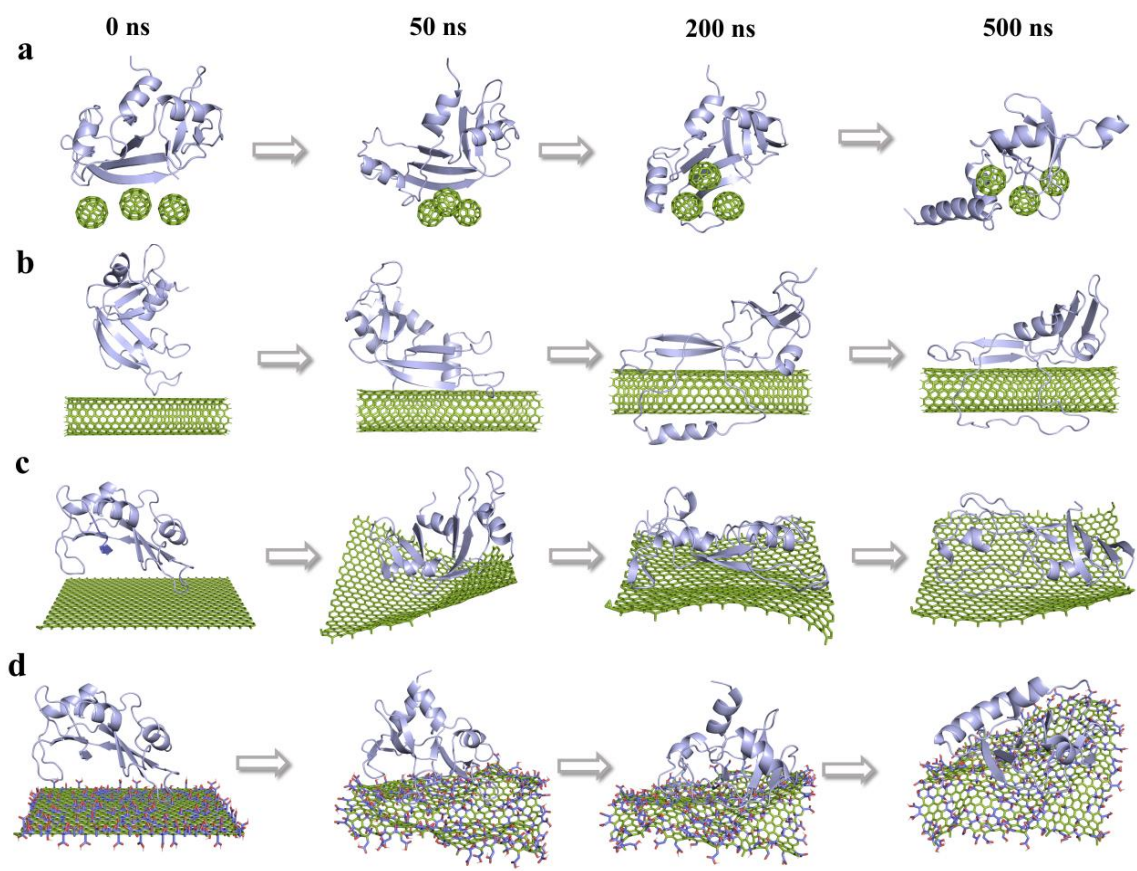

**Fig. 2. The absorption process of RNase A onto (a)** C60, **(b)** SWCNT, **(c)** PG and **(d)** GO.

In order to gain a comprehensive understanding of the interaction mechanisms between GBNMs and protein at the molecular level, molecular dynamics simulations were carried out to investigate the interaction behavior at the interface between RNase A and GBNMs. These simulations employed a customized force field tailored for GBNMs referring to the reported literature (Muraru et al. 2020; Zeng et al. 2015; Vögele et al. 2018; Cao et al. 2009). Absorption behavior of protein onto GBNMs were observed and analyzed, and thus making the changes in protein conformation readily apparent and visually discernible during the simulation process. Fig. 2 and Fig. S1 displayed the initial and absorbed snapshots of different GBNMs versus RNase A and ExoIII with at various time points, respectively. The absorption process can be divided into three stages: in the first stage, protein was robustly attracted and gradually approached to GBNMs (0-50 ns), and maintained most of its secondary and tertiary structures with minor fluctuations; in the second stage, protein was strongly anchored on the GBNMs surface (200 ns) with clearly visible structural changes; in the third stage, most part of the residues attached to the GBNMs surface and in risk of protein structural collapsing (500 ns).

The characterization of absorption process including all-atom root mean square deviation (RMSD), distance, contact area and energy during simulation to specifically illustrate this kinetic adsorption process (Fig. 3). The initial state of proteins was around 2 nm from the center of mass of protein to the surface of GBNMs. After minimization, heat and equilibrium, production run simulation was carried out. For RNase A, the structure was rapidly absorbed onto surface of GBNMs during the production simulation. Fig. 3a demonstrated that all-atom RMSD fluctuations of RNase A stemmed from PG, SWCNT and C60 can reach up to an average of 7 Å while GO shows fluctuations of around 2 Å. The distance (Fig. 3b) between RNase A and PG, C60 and SWCNT was rapidly decreased owing to the robust absorption, which may company with the collapse of the RNase A structure. While GO exhibited comparatively weaker absorption compared to the other three materials, with RNase A slowly being drawn to its surface. Moreover, the contact surface (Fig. 3c) between RNase A and GBNMs rapidly increased during simulation time of 10-150 ns, and RNase A was entirely anchored onto nanomaterials since 300 ns. The contact area of GO, PG and SWCNT was larger than that of spherical C60. Additionally, the non-covalent adsorption energy including Van der waals force and electrostatic energy change was almost constant after 250 ns regarding to energy profile for four GBNMs (Fig. S2a). The comprehensive simulation outcomes indicated that GO exerted the minimal impact on the conformational integrity of RNase A. The strong interactions with PG, SWCNT, and C60 significantly influenced atomic movements and positional changes, potentially leading to more pronounced alterations in the secondary structure of RNase A. However, among these, C60 displayed a marginally lower interaction energy compared to PG and SWCNT. This could be attributed to C60's relatively smaller particle size and the correspondingly reduced contact area with the proteins.

For ExoIII, another candidate protein, exhibited minor fluctuations during simulation with all four GBNMs comparing with RNase A, ranging from 2-4 Å as depicted in Fig. 3d. During the absorption behaviour between GBNMs and ExoIII, PG maintained the stronger absorption capacity (Fig. 3e). This may be due to the larger contact area with the planar material PG compared to cylindrical SWCNT and spherical C60, when absence of a collapse in protein structure. Notably, during simulation time ranging from 10-150 ns, the overall contact area remained in a slow increasing kinetics, which is significantly distinguished from that of RNase A (Fig. 3f). A larger structure of ExoIII (31kDa) resulted in a larger contact area compared with RNase A (13.7kDa). Similarly, the overall interacting energy of four GBNMs (Fig. S2b) was stronger than that of RNase A. Taking a close look in graphene, the heat map showed that for RNase A and, the content of α-helix and β-strand, especially β-strand decreased gradually and significantly with time in simulation (Fig. S3a-b). Such significant disruption of secondary structure from 'planar' π-π interaction may potentially lead to dysfunction of RNase A activity.

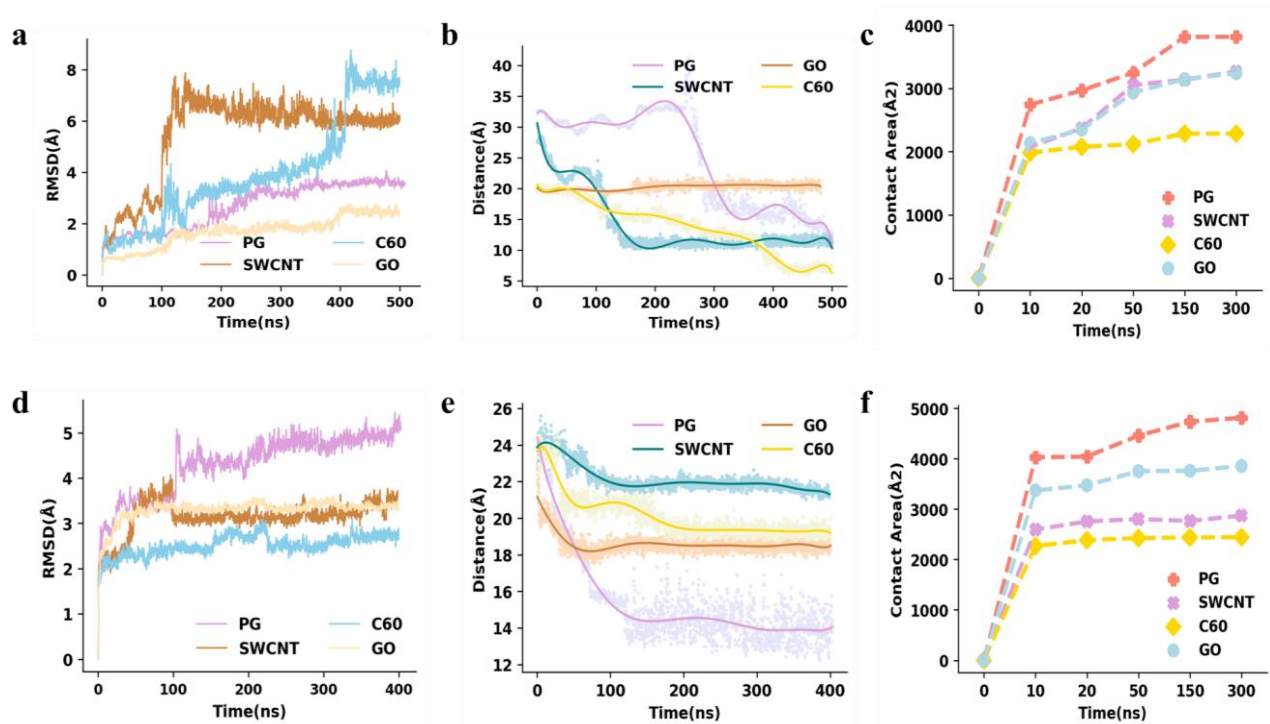

**Fig. 3. Molecular dynamics simulation between proteins and GBNMs. (a-c)** was for RNase A and **(d-f)** was for ExoIII. The all atom RMSD fluctuation of **(a,d)**, the distance **(b, e)**, contact surface area **(c, f)** between proteins and GBNMs during simulation.

*3.2 Protein activity simulations*

From the integrated perspective, after 500 ns simulation, the conformation change of RNase A was evident (Fig. S4a), compared with the original structure, PG, SWCNT and C60 had relatively large variation in secondary structure, mostly decrease in α-helix and β-sheet (Fig. 4a). On the contrary, GO had a minor influence on RNase A conformation, suggesting it may retain its activity. This phenomenon can likely be attributed by π- π stacking interaction between the amino acids with 'planar' side chains forming β-strands, including TYR73, PHE46 and TYR97 with sp$^2$ hybridization triggered honeycomb crystal lattice in graphene (Fig. S5). Unlike PG, SWCNT and C60, GO contains oxygen functional groups attached to the carbon lattice, resulting in a more hydrophilic and negatively charged surface. The polar surface can engage in hydrogen bonding and electrostatic interactions, which will attract protein absorbed onto surface but not cause significant changes to protein secondary structure. Thus, the "planar" π-π interactions from PG, SWCNT and C60 is indicated to be more potent than electrostatic and hydrogen bonding from GO in disrupting protein secondary structures. The initial orientation of protein also has impact on its adsorption behavior, for interface containing most helical structures, it is more likely that protein will be rapidly adsorbed onto nanomaterial surface. As mentioned in previous research (Wei et al. 2019) that the strong interaction may result in α-helical conformation change. The simulation results suggested that it can also lead to the distortion and deformation of β-strands, contributing to the finally structural changes.

In addition, to further analyze the bioactivity of the interacted proteins, the enzyme active site of RNase A underwent close inspection, as shown in Fig. 4b. The cavity enclosed by HIS12 and HIS119 is the location where substrate binding and catalysis process takes place. However, upon interaction with GBNNs, the site experienced disruption due to displacement of these two residues, signifying a loss of activity and functionalization. Furthermore, molecular docking was performed to examine if the alternation of tertiary structure will maintain the affinity with ligand substrates. For molecular docking, the input comprised both the original conformation and the conformation with the highest RMSD for comparison, the output affinity score directly reflected the bioactivity of target protein. In the case of RNase A, five ligand molecules, all containing a phosphate group which served as the target for hydrolysis, were selected for docking. The results were in consistent with the predictions, GO interacted RNase A obtained similar affinity score as the original structure, while other nanomaterials decreased the affinity by displacing the ligand-favored residues (Fig. 4c).

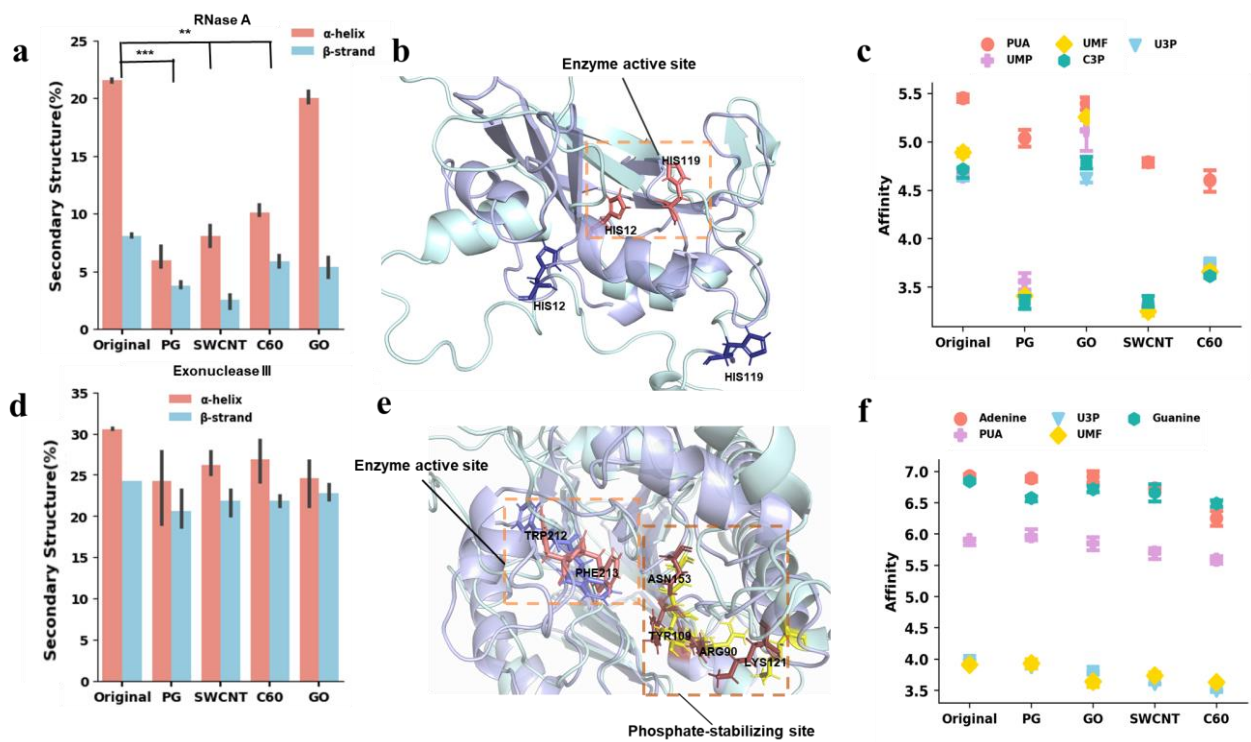

**Fig. 4. Prediction of protein bioactivity**. **(a)** Secondary structure contents of RNase A before and after interacting with GBNMs, **(b)** Enzyme active site (HIS12, HIS119) of RNase A and place displacement of them after interaction, **(c)** Affinity score of RNase A docking with five compounds(PUA, UMP, UMF, U3P, C3P), **(d)** Secondary structure content of ExoIII before and after interacting with GBNMs, **(e)** Enzyme active site of ExoIII (TRP212, PHE213) and phosphate stabilizing site before and after the interaction, **(f)** ExoIII with Adenine, U3P, Guanine, PUA and UMF containing phosphate group before and after interacting with GBNMs. Error bars represent the standard deviation of three independent measurements.

From the integrated perspective, after 500 ns simulation, the conformation changes of ExoIII interacting with different GBNMs are very subtle (Fig. S4b) compared with the original structure. There was only minimal change of secondary structure content in ExoIII compared with RNase A, implying the preservation of its bioactivity following interacting with GBNMs (Fig. 4d). On the other hand, the enzyme active site and phosphate-stabilizing site (Lee et al. 2022) of ExoIII remained intact, maintaining a good alignment with the original structure after interacting with GBNMs (Fig. 4e). Subsequently, with the same strategy as RNase A for target protein, additional base molecules of different adenine and guanine were selected as docking molecules of ExoIII to illustrate the activity of the protein. And the resulting affinity score for all the GBNMs remained with high similarity (Fig. 4f).

3.3 Characterization of absorption behavior

The simulation results of RNase A and ExoIII were also confirmed by CD spectra (Fig. S6), as known that the α-helix typically exhibits signal peaks at approximately 190 nm, 222 nm, and 208 nm, while the β-sheet may display characteristic signal peaks within the 185-200 nm range and around 218 nm (Greenfield 2006). A shift in the CD peaks indicated alterations in the protein's secondary structure. After calculation of α-helix and β-sheet composition separately and compared with original conformation, it was found that β-sheet varied massively for PG, SWCNT and C60-interacting RNase A, and showed a noticeable decrease in the composition (Fig. S7a, Table S2). In contrast, GO exhibit minimal diversity in the overall secondary structure, with only a slight increase in the α-helix composition which is unlikely to influence the function of the protein. These results were in consistent with the simulation predictions. For ExoIII, all protein-GBNMs exhibited similar tendency and curve shape as the original protein (Fig. S7b, Table S3). Consequently, the secondary structure variable was significantly smaller than that of RNase A while interaction with GBNMs, which is also in consistent of the simulation prediction.

In addition, when considering the interaction of protein with nanomaterials, DLS and AFM characterization of protein-nanomaterial binding results (Fig. 5a-d) revealed the effective absorption of proteins onto the GBNMs surface. Herein, RNase A was selected for illustration. Due to the challenge posed by the extremely small size of C60, with a diameter of less than 3 nm, thus dynamic light scattering (DLS) was employed to analyze the diameter of C60 both before and after combination with protein. As illustrated in (Fig. 5a), the inherent aggregation tendency of C60 particles, preventing complete disperse in water, led to an initial diameter around 18 nm. Upon formation of the protein-C60 complex, the diameter increased to around 28 nm, which provided evidence that RNase A adhering to the surface of C60. Fig. 5b-c exhibited PG and GO exhibited an evident sheet-like structure with an initial thickness ranging from 1 to 3 nm. After the interaction with protein, the thickness height increased up to around 8-9 nm, suggesting proteins have anchored onto the surface. Similarly, the initial diameter of SWCNT (Fig. 5d) was about 2 nm, and when combined with RNase A, the thickness increases significantly.

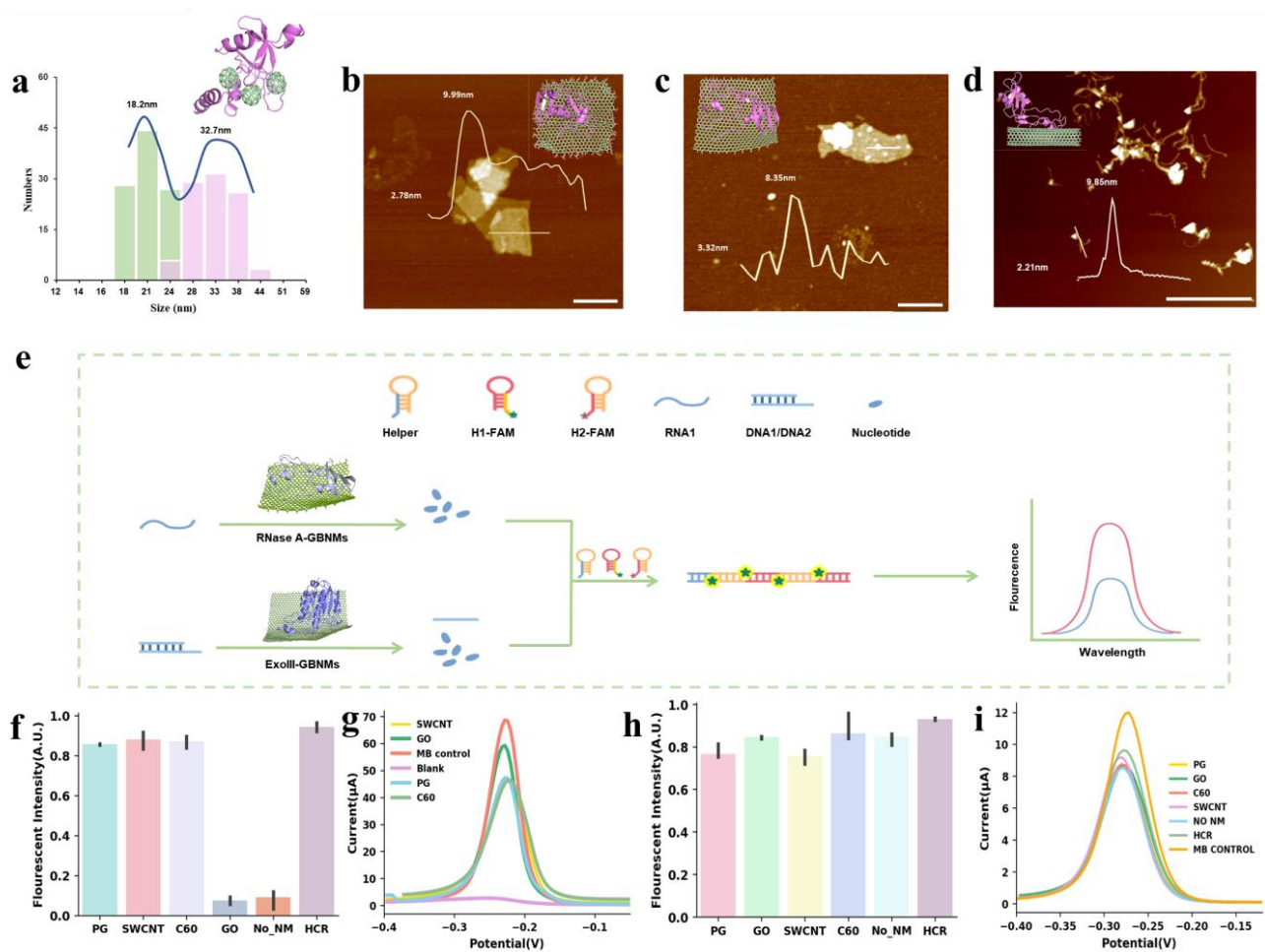

**Fig. 5. Protein characterization of absorption behavior and bioactivity**. **(a)** The DLS result of C60 nanoparticles. **(b-d)** AFM images of protein absorption on SWCNT, PG and GO. Scale bar: *400 nm*. **(e)** Schematic drawing of fluorescence measurement to assess the activity of RNase A and ExoIII by checking the existence of HCR. Fluorescence signals for detecting the exisition of HCR in the case of **(f)** RNase A and **(g)** ExoIII. DPV signals for the existence of HCR in the case of **(h)** RNase A and **(i)** ExoIII.

*3.4 Experimental examination of protein bioactivity*

According to the simulation results, it is suggested that GO was an excellent substance that can retain RNase A activity with intact protein structure and all four GBNMs were appropriate for ExoIII. Herein, the fluorescence and electrochemical experimental validation was employed. Hybridization chain reaction (HCR), an isothermal amplification technique (Zhu et al. 2021), was utilized to enhance both fluorescence and electrochemical signals. In the absence of RNA1, the mixture of helper, H1 and H2 remained in single-strand hairpin status (Fig. S8a, lane 5); when RNA1 was present (Fig. S8a, lane 6), various stripes with different molecular weight appeared, indicating the successful assembly of HCR structure. On the contrary, when RNase A was added (Fig. S8a, lane 7), the various strips were not observed, showing that hydrolyzed RNA1 by RNase A resulted in the incapability of assembling HCR structure. Therefore, the HCR design could be used to identify the difference of RNase A bioactivity brought by GBNMs absorption.

As exhibited in Fig. 5e, RNA would be hydrolyzed to nucleotide when RNase A-GBNMs bioconjugates stained bioactivity, therefore FAM- H1 and FAM- H2 will be absorbed at the surface of nanomaterials and showed lower fluorescence intensity (Liu et al. 2013). If the structure of RNase A was disrupted by GBNMs, RNA would activate hairpin DNA, triggering the HCR reaction. The long-stranded HCR duplex will be detached from nanomaterial surface, resulting in an increased fluorescence signal. The florescence results (Fig. 5f) illustrated that the successful assembly of HCR structure for PG, SWCNT and C60-RNase A complex with high luminescence. This outcome suggested that the dysfunction of RNase A led to the persistence of RNA. In contrast, GO was incapable of disrupting RNase A activity, resulting in the similar florescence intensity to the control sample which didn't involve the addition of any nanomaterials, implying that the RNA has lost its activity due to hydrolysis by RNase A. Moreover, we also employed electrochemical sensing method (Fig. S9) to detect DNA hybridization with methylene blue. Duplex HCR products would absorb a large amount of MB (an electrochemical indicator), leading to a weaker electrochemical signal. On the contrary, singled-stranded DNA when without the existence of the HCR reaction, a stronger electrochemical signal will be observed. DPV measurements (Fig. 5g) displayed a consistent tendency. In comparison to MB control, PG, SWCNT, C60-RNase A complexes which facilitated the formation of HCR assembly, exhibited a weaker and similar current. Conversely, GO-RNase A complex resulted in a larger current. The DPV measurements exhibited same tendency with fluorescent intensity results, which is consistent with the results of computer simulation.

In the case of ExoIII, the duplex DNA hybridized by DNA1 and DNA2 would be hydrolyzed by ExoIII and released the single-strand target DNA1, resulting in the occurrence of HCR reaction. This process was confirmed by PAGE (Fig. S8b), after addition of ExoIII, the formed duplex DNA (Fig. S8b, lanes 3-4) was hydrolyzed and discharged target DNA1. With the input of helper, H1 and H2 after deactivation of ExoIII, series strips were shown (Fig. S8b, lane 5), indicating the successful assembly of HCR structure. Additionally, florescence examination observed high luminescence for all GBNMs in Fig. 5h, representing the existence of HCR structure and the successful cleavage by ExoIII, which indicated the preserved bioactivity of ExoIII after interacting with all four GBNMs. Consisting with the fluorescence results, DPV exhibited in the similar amplitudes, implying there was no distinguished change of the bioactivity after interaction with GBNMs (Fig. 5i), all the experiment results successfully validated our simulation prediction.

In addition, the bioactivity can be quantified and correlated with the fluorescence intensity, DPV signals and averaged docking affinity with chemical compounds, the prediction showed in great alignment with simulation-florescent, simulation-DPV and DPV-florescent, all three pairs revealed a linear correlation with good confidence interval (Fig. 6a-b). Due to the reason that the measurement and affinity score of RNases A were distinguished, they were reasonably well-fitted, reaching a correlation coefficient of 0.96. On the other hand, the similar level of ExoIII bioactivity for all the nanomaterials resulted in a trustworthy correlation of 0.82. However, the interaction between ExoIII and GBNMs consistently maintained a stable affinity, posing challenges in precisely quantifying the extent of bioactivity.

Furthermore, in order to reconfirm the methodology and pinpoint the optimal structure snapshot to obtain the most accurate affinity for assessment of bioactivity of proteins for molecular docking, different frames from molecular dynamics simulation were extracted and compared, including the frame with largest RMSD (Fig. 6c-d) and the last frame (Fig. 6e-f). The correlation efficient of last frame (0.63 for RNase A, 0.1 for ExoIII) was significantly lower than that of the largest RMSD (0.98 for RNase A, 0.82 for ExoIII). Thererfore, the proteins snapshot structure with largest RMSD was chose as candidate for docking. Relative to RNase A, the correlation efficient of ExoIII was slightly lower because the affinity of ExoIII during interacting with GBNMs remained quite consistent, making it difficult to accurately quantify the level of bioactivity.

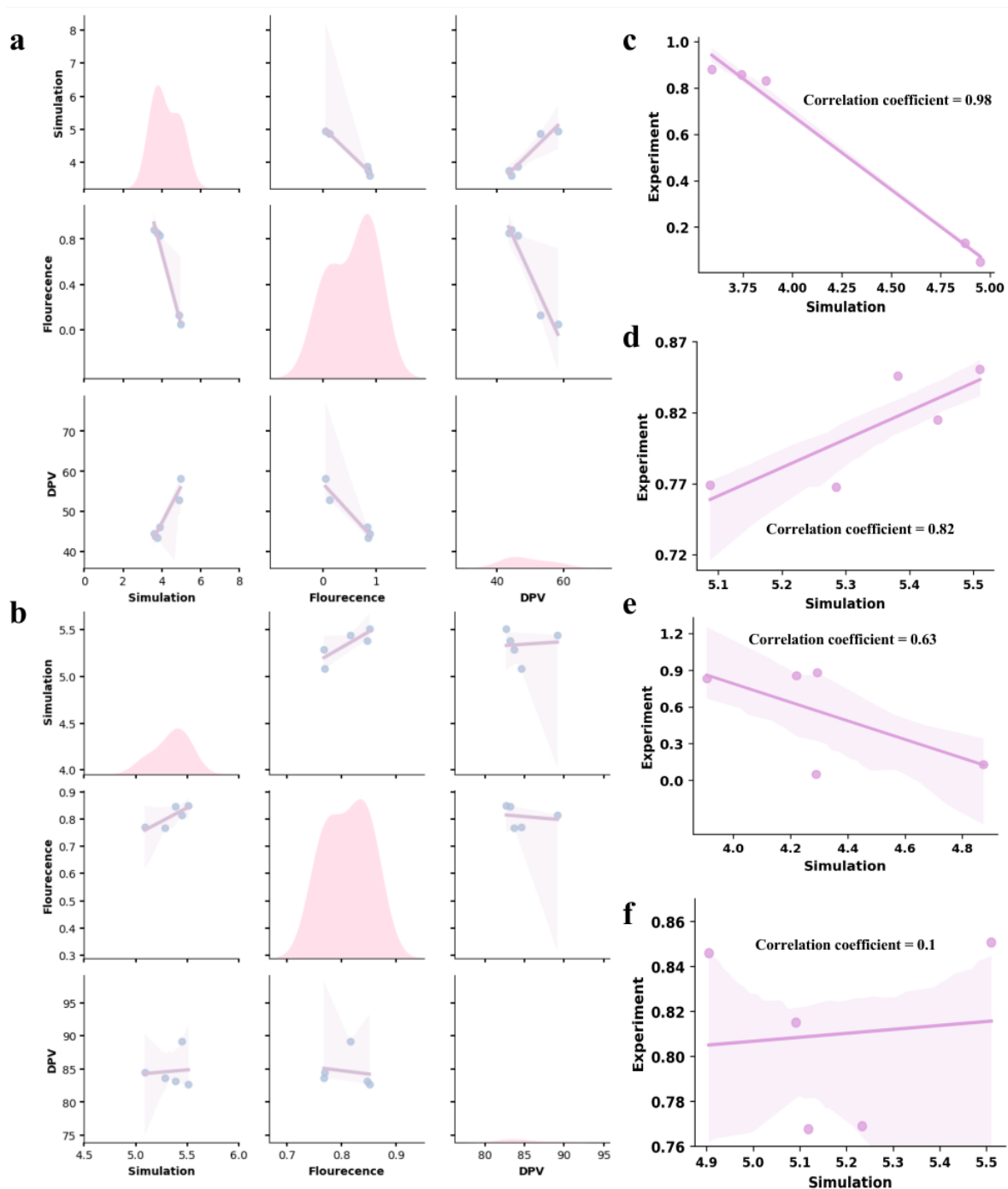

**Fig. 6. Experiment and simulation prediction alignment.** Alignment of simulation prediction with experiment results of **(a)** RNase A and **(b)** ExoIII. Fluorescence experiment and simulation prediction alignment with different frames of trajectory with the **(c-d)** largest RMSD frame during the simulation, with the **(e-f)** last frame during simulation.

*3.5 Cytotoxicity experiments of GBNMs*

Up to this point, both simulation and experimental outcomes have demonstrated the optimal choice for GBNMs *in vitro*, providing a satisfactory basis for RNase A and ExoIII based biosensor design. For biomedical applications *in vivo*, such as drug delivery, vaccine carrier etc. Cytotoxicity is a crucial element to assessment. Therefore, in vivo experiments were carried out to evaluate weather PG, GO, SWCNT and C60 have cytotoxicity towards A549 cells, the viability of A549 cells was tested after 24 h incubation, with treatments involving different concentrations ranging from 1 μg/ml to 50 μg/ml for each material. As shown in Fig. 7a, GO demonstrated the highest cell viability across all concentrations, even when exposed to high concentrations (50 μg/ml), the viability remained above 70%. In contrast, SWCNT and C60 showed deleterious effect on cell viability at concentrations exceeding 20 μg/ml. Notably, at a concentration of 50 μg/ml, C60 exhibited extremely detrimental effect on A549 cells, resulting in a viability less than 20%. On the other hand, PG exhibited reduced toxicity to cells compared with SWCNT and C60, still maintained relatively good viability (> 65%) at high concentrations.

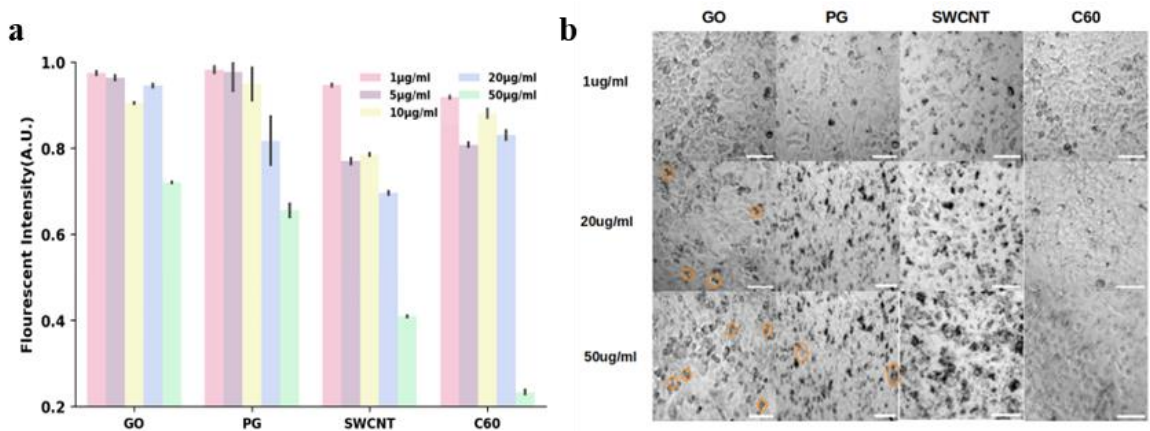

**Fig. 7. Cytotoxicity of GBNMs**. **(a)** Cell viability of A549 cells after incubation with different concentrations of PG, GO, SWCNT and C60. **(b)** Microscopic recordings of cell status with four nanomaterials, scale bar: *100 μm*.

Microscopic observations further supported these findings. As depicted in Fig. 7b, cells still displayed division behavior (marked by orange circle) even under high concentrations, implying their sustained high activity status. In the case of PG, several dead cells were observable at 50 μg/ml, while for SWCNT, the cells were covered with dark nanotube particles, and it was evident that cells were mostly dead after treated with 50 μg/ml C60 solutions. Overall, cytotoxicity assessment results illustrated that GO has the least cytotoxicity to A549 cells, while SWCNT and C60 demonstrated a significant toxicity. Apparently GO is the most bio-compatible material among the four graphite carbon family materials, implying its potentiality of a promising candidate for use in monotherapies or drug delivery in the future. For instance, currently most mRNA vaccines are attached and coated with lipid particles, a drawback of the mRNA-lipid nanoparticle vaccines is that they have to be stored at ultra-low temperature, otherwise they will be highly unstable. Therefore, it is significant to develop another mRNA vaccine payload (Yin et al. 2021) that can immobilize mRNA but also does not affect the activity of it, in addition, harmless to human physiological system. Graphene oxide might have such potential, it can be effective building blocks for different physical and chemical properties, like GO hydrogel (Yi et al. 2020). Furthermore, their additional functional groups can be modified with other micro/macro molecules, to realize extensive applications such as drug delivery, tissue scaffolds and biosensors.

## 4. Conclusion

In this study, the interaction mechanism and impact on protein bioactivity was investigated by numerical method, primarily consolidated molecular dynamics simulation and molecular docking-based techniques, along with the aid of comprehensive experimental methods have been applied to verify the protein bioactivity. The investigation encompasses four key aspects: absorption behavior, secondary structure alterations, protein activity, and cytotoxicity. In general, simulation results suggested that all graphene-like nanomaterials had strong protein absorption, especially dominated by strong π-π interaction originating from aromatic residues. Additionally, in some case, such interaction may result in the disruption of protein secondary structure, potentially leading to reduced protein activity. This finding was further validated by florescence assay and electrochemical experiments, which employed the mechanism of HCR. Both experiment and simulation results suggested that PG, SWCNT and C60 may induce RNase A dysfunction by interfering its conformation, GO demonstrates the capability to maintain RNase A bioactivity alongside effective absorption. As well as all four GBNMs held good compatibility with ExoIII. Our study showed an easy-to-operate simulation modelling to evaluate the effects of graphene-based nanomaterials on protein activity, with a prediction accuracy reaching up to 0.98. It also explains protein inactivation at the molecular level, along with the explanation for the differences in biosensor performance. The method can also be extended to other nanomaterials and proteins, which was meaningful for protein-related biosensor.